\documentclass{article} 
\usepackage{nips12submit_e,times}

\RequirePackage{amsthm,amsmath,amsbsy,amssymb,bm,color,graphicx,epstopdf,subfigure}

\title{Order Statistics of Observed Network Degrees}

\author{Sofia C. Olhede and Patrick J. Wolfe\\
Departments of Statistical Science and Computer Science\\
University College London\\
London WC1E 6BT, UK \\
\texttt{\{sofia, patrick\}@stats.ucl.ac.uk} \\
}

\nipsfinalcopy

\begin{document}

\maketitle

\begin{abstract}
This article discusses the properties of extremes of degree sequences calculated from network data.  We introduce the notion of a normalized degree, in order to permit a comparison of degree sequences between networks with differing numbers of nodes. We model each normalized degree as a bounded continuous random variable, and determine the properties of the ordered $k$-maxima and minima of the normalized network degrees when they comprise a random sample from a Beta distribution. In this setting, their means and variances take a simplified form given by their ordering, and we discuss the relation of these quantities to other prescribed decays such as power laws. We verify the derived properties from simulated sets of normalized degrees, and discuss possible extensions to more flexible classes of distributions.
\end{abstract}

\section{Introduction}\label{sec:introduction}
Networks are ubiquitous as models of relational data in science and engineering; their structure is described via linkages between the set of nodes of the network. Linkages are commonly assessed in terms of which nodes they connect; for example, important nodes may connect to many other nodes, while less important nodes may make only a few connections.  In this article, we study the propensity of the most important nodes for linkage, as assessed in terms of the degree sequence of the overall set of network nodes.  In particular, we study the properties of the extreme degrees, and discuss their moments for a flexible class of (normalized) random degrees based on the Beta distribution. We also investigate their properties via simulations, and show how the derived properties relate to observed extreme degrees.

Each linkage between two nodes can either be present or absent, and so is modelled by a Bernoulli random variable (independent of all others) taking the value zero or one \cite{chung2006complex,bollobas2001random}. We collect the links between any two nodes in a matrix $\bm{A}$, known as the {\em adjacency matrix}, which by necessity is symmetric.  The sum of all linkages associated with a single node $i$ gives the {\em degree} of that node; in the simplest sense the degree $d_i$ indicates the importance of node $i$ to the network \cite{Durrett}. Starting from the matrix $\bm{A}$, this means summing the matrix entries over row $i$ to form the vector of degrees $\bm{d} = \bm{A} \bm{1}$, where $\bm{1}$ is the vector of all ones.
Common practice is to study the full set of degrees $\bm{d}$ for a network, in order to identify the most important (or least important) nodes; i.e., in this context, those that have especially extreme degrees.

We study the properties of extremes of degree sequences, to understand the degree of diversity that can be achieved simply by the act of ordering an identically distributed sequence of random variables corresponding to observed degrees. To remove an arbitrary dependence on the size $n$ of the network, we rescale each degree by the total number of nodes $n$ to yield a {\em normalized degree}, or proportion $\pi_i=d_i/n$. This normalized degree can take a value between zero and one. We therefore model the normalized degrees according to a distribution supported on $[0,1]$. The normalized degrees can also only take quantized values of proportions (i.e., the proportion of other nodes that are connected to node $i$), but for sufficiently large networks, approximating their distribution with that of a continuous random variable does not significantly alter the properties of the rescaled degrees. We also note that network degrees are mildly correlated under the model specified above, as they are constructed from the same random matrix and each degree pair shares one potential linkage.  However, as the magnitude of this correlation decreases with increasing sample sizes, this effect can safely be ignored as $n$ grows large, so that subsequently we treat these random variables as a random sample from a given distribution.

In this setting we study the extremes of the normalized degrees.  Rescaled extreme values can, if a limiting distribution exists, take one of three possible forms \cite{lindgren1987extreme,Leadbetter}. The choice of rescaling has a pivotal role on the properties of the observed extremes. Starting from a bounded density, such as a $\operatorname{Uniform}(0,\theta)$ or (more generally) a $\operatorname{Beta}(a,b)$ density, we will obtain the (reversed) Weibull distribution \cite{lindgren1987extreme,Leadbetter}. We show in this case how the expected values of the $k$ largest (or smallest) variates take the form of a simple curve in terms of their size ordering $k$, creating an \emph{apparent} hierarchy of the degree sequence.  This effect occurs as a consequence of the behavior of order statistics, \emph{despite} the degrees corresponding to a random sample of (in point of fact, approximately) independent and identically distributed variates. Similarly, the variances of the $k$ maxima take a simple form depending on their ordering. These properties depend on the parameters of the distribution under consideration.  The parameter governing the Beta distribution near its right tail dominates the expectation and variance of the maxima, and that governing its left tail dominates those of the minima.

We conclude by briefly investigating the properties of simulated normalized degree sequences. We show for certain choices of Beta parameters $a$ and $b$ how the expected normalized degrees exhibit the derived properties over a full range of ordered indices. We then compare these means with a popular choice of mean decay---that of a \emph{power law} \cite{Durrett}---and show that the expected ordered degrees can be approximately described in terms of a power law, if we allow for the form of a shifted power law as introduced in \cite{chung2003spectra}.

\section{Summary of main results}
\label{sec:orderStatResults}

Our main results concerning order statistics of normalized observed degree sequences are as follows.

Suppose $\pi_{(1)} \geq \pi_{(2)} \geq \cdots \geq \pi_{(n)}$  are ordered elements from a $\operatorname{Beta}(a,b)$-distributed random sample of size $n$ with parameters $a,b > 0$, intended to model ordered proportions of degrees divided by the number $n$ of nodes in the network.  Then:

\begin{enumerate}

\item As $n \rightarrow \infty$, the distribution of $n^{1/b}(1 - \pi_{(k)})$ for fixed $k$ converges to a generalized Gamma distribution with scale parameter $(\beta(a,b)b)^{1/b}$ and shape parameters $(k,b)$, corresponding to the Weibull distribution for $k=1$.

\item When $n$ is large and $k$ is fixed, the first two moments of $\pi_{(k)}$ can be expressed as follows:
\begin{align*}
\operatorname{\mathbb{E}} \pi_{(k)} & \asymp 1 - \frac{\Gamma(1/b)}{\beta(k,1/b)} \left(\frac{\beta(a,b)b}{n}\right)^{1/b} \sim 1 - \left(\beta(a,b)b \, \frac{k}{n}\right)^{1/b},
\\ \operatorname{Var}\left(\pi_{(k)} \right) & \asymp \left[ 1 \! - \! \operatorname{\mathbb{E}} \pi_{(k)} \right]^2 \left[ \frac{\beta(k,1/b)}{\beta(k + 1/b,1/b)} - 1 \right] \sim \left[ 1 \! - \! \operatorname{\mathbb{E}} \pi_{(k)} \right]^2 \left[ \left(\frac{k + 1/b}{k}\right)^{1/b} \!\!- 1 \right].
\end{align*}
The latter approximation to $\operatorname{\mathbb{E}} \pi_{(k)}$ is close to the former for $a,b>0$, while the latter approximation to $\operatorname{Var}\left(\pi_{(k)} \right)$ is close to the former for $b\geq 1$, or $kb \gg 1$ when $b<1$.

\item Analogous asymptotic results hold for $\pi_{(1)} \leq \pi_{(2)} \leq \cdots \leq \pi_{(n)}$, with $\beta(a,b)b$ replaced by $\beta(a,b)a$ and $1/b$ replaced by $1/a$ in the discussion and equations above.  To wit, for the $j$th smallest order statistic $\pi_{(j)}$, the distribution of $n^{1/b}\pi_{(j)}$ for fixed $j$ approaches a generalized Gamma distribution with scale parameter $(\beta(a,b)a)^{1/a}$ and shape parameters $(j,a)$, and thus $\operatorname{\mathbb{E}} \pi_{(j)} \sim \left(\beta(a,b)a \cdot j/n\right)^{1/a}$ as $n \rightarrow \infty$.

\end{enumerate}

For $a=1$, the asymptotic expression in $k$ converges quickly in $n$ and holds throughout the sequence, while for $b=1$ that in $j$ behaves similarly.  These expressions are closest to being exact in the case of the uniform distribution ($a=b=1$), whereupon (as we show below) the expectation of the $j$th smallest sample element is given by $\operatorname{\mathbb{E}} \pi_{(j)} = j/(n+1)$, or more generally $\theta j/(n+1)$ when values are distributed uniformly on the interval $[0,\theta]$.

\section{Exact and limiting distributions of order statistics for normalized degrees}

We are interested in the following question: if $\pi_1, \ldots \pi_n$ comprise a random sample from a given density $f(\pi)$ whose support is contained in $[0,1]$, and whose tail behavior is specified, then what decay do the \emph{ordered} sample values exhibit as they approach zero or one?  Answering this question allows us to characterize network nodes that have especially extreme degrees, as described above.

\subsection{The special case of $\operatorname{Uniform}(0,\theta)$ variates}

We shall develop our understanding of the extreme values of $\pi_i$ in a series of steps, starting first from simple uniform random variables, and then building up our intuition to more complicated distributions. We first recall some basic results for order statistics.  We index sample elements from smallest to largest as $\pi_{(1)}, \ldots \pi_{(n)}$, with $\pi_{(j)}$ being the $j$th smallest, recalling
\begin{align*}
F_{X_{(j)}}(x) & = \sum_{i=j}^n \binom{n}{i} [F(x)]^i [1-F(x)]^{n-i},
\\ f_{X_{(j)}}(x) & = \binom{n}{j} j  [F(x)]^{j-1} [1-F(x)]^{n-j} f(x).
\end{align*}
From these expressions we can compute $\operatorname{\mathbb{E}} X_{(j)}$ directly, or via the identity $\operatorname{\mathbb{E}} X_{(j)} = \int_{0}^{\infty} (1-F_{X_{(j)}}(x)) \, dx$ if $\operatorname{supp} f(x) \subseteq \mathbb{R}^+$.  It follows that the expected value of the $j$th smallest element $\pi_{(j)}$ from a $\operatorname{Uniform}(0,\theta)$ random sample is hence straightforwardly obtained as
$$
\operatorname{\mathbb{E}} \pi_{(j)} = \theta \frac{j}{n+1}, \quad j = 1, 2, \ldots, n.
$$

\subsection{$\operatorname{Beta}(a,b)$ order statistics and extreme value theory}

The $\operatorname{Uniform}(0,\theta)$ case considered above is interesting because we see clear structure in terms of the ordering of the random sample. It now becomes sensible to extend such a result to distributions that show greater preferential weighting to parts of their range.  We recall that the $\operatorname{Uniform}(0,1)$ distribution is a special case of the $\operatorname{Beta}(a,b)$ distribution for $a = b = 1$.  Obtaining exact results for the general case of $\operatorname{Beta}(a,b)$ variates is possible for integral $(a,b)$ by way of recursions involving higher-order moments of a $\operatorname{Beta}(a,b)$ variable~\cite{nadarajah2008explicit, nadarajah2008review, thomas2008recurrence}.
We may also appeal to extreme value theory by way of the cumulative distribution function (CDF) of the Beta density.  Continuing in this direction, we now investigate the limiting behavior (in $n$) of order statistics of a random sample $\pi_1 \ldots \pi_n$ from the $\operatorname{Beta}(a,b)$ distribution.

Choosing constants $a_n = n^{1/b}$ and $b_n = 1$ with foresight, we will show that the law of the maximal order statistic, suitably transformed to $u = a_n(\pi_{(n)} - b_n)$, converges to a nondegenerate distribution function $G(u)$.  We do this as follows.

\begin{enumerate}
\item First, we observe that the CDF $F(x)$ of a $\operatorname{Beta}(a,b)$ random variable is given by the regularized incomplete Beta function
$$
I_x(a,b) = \beta(a,b)^{-1} \int_0^x t^{a-1} (1-t)^{b-1} \, dt, \quad 0 \leq x \leq 1,
$$
and so we conclude that the law of $\max_i \{\pi_i \overset{\mathit{iid}}{\sim} \operatorname{Beta}(a,b) \}_{i=1}^n$ is given by $[I_x(a,b)]^n$.

\item Second, we must show that the CDF of $u$ goes to some limiting law $G(u)$ as $n \rightarrow \infty$, and so we must verify that $\lim_{n \rightarrow \infty} [I_x(a,b)]^n$ exists.
\end{enumerate}

We proceed as follows, first considering the case when $a$ and $b$ are positive integers.

\subsubsection{Extreme values of $\operatorname{Beta}(a,b)$, with $a,b \in \mathbb{N}$}

Observe that for $a,b \in \mathbb{N}$, the CDF $F(x)$ of a $\operatorname{Beta}(a,b)$ random variable reduces to
$$
I_x(a,b) = \sum_{i=a}^{a+b-1} \binom{a+b-1}{i} x^i (1-x)^{a+b-1-i}, \quad 0 \leq x \leq 1, \quad a,b \in \mathbb{N},
$$
and so to simplify calculations we shall commence
by assuming that the parameters are integer valued.  Observe that by the binomial theorem,
$$
1 = I_x(a,b) + \sum_{i=0}^{a-1} \binom{a+b-1}{i} x^i (1-x)^{a+b-1-i}, \quad a,b \in \mathbb{N},
$$
and so
$$
\left[I_x(a,b)\right]^n = \left[1 - \sum_{i=0}^{a-1} \binom{a+b-1}{i} x^i (1-x)^{a+b-1-i} \right]^n, \quad a,b \in \mathbb{N}.
$$

Since we are concerned with $\max_i \{X_i \overset{\mathit{iid}}{\sim} \operatorname{Beta}(a,b) \}_{i=1}^n$, we expand $x$ near 1 in $-u$ as follows:
$$
x = 1 + \frac{u}{n^{1/b}},
$$
which corresponds to the choice $a_n = n^{1/b}, b_n = 1$ with respect to the limiting behavior we wish to explore.  We define the function $G_n(u)$ as $[I_{1+u n^{-1/b}}(a,b)]^n$, and
detailed calculations in Appendix A yield that $G_n(u)$ for $a,b \in \mathbb{N}$ is given by
\begin{equation}
\label{Gnu}
G_n(u) = \left[1 - \frac{1}{\beta(a,b)b} \left(1 + \frac{u}{n^{1/b}} \right)^{a-1}
\left( \frac{-u}{n^{1/b}} \right)^{b} \left\{ 1 + f_{ab}(u,n) \right\} \right]^n,
\end{equation}
where $| f_{ab}(u,n) | \leq   \left[(a-1)/(b+1)\right] (u / n^{1/b}) \left(1
+ u/n^{1/b}\right)^{1-a} \mathbb{I}(a \geq 2)$. This yields the CDF of the rescaled and shifted random variable $n^{1/b}(x-1)$, and as we shall see, additional insight can be determined from~\eqref{Gnu}.

\subsection{The limiting form of extreme values}

Now observe that as $n$ grows large, for fixed $u$ we have that $G_n(u)$ approaches $[1-(\beta(a,b)b)^{-1}(-u)^b/n]^n$.  Thus for fixed $u$ and as $n \rightarrow \infty$, we obtain the limiting form
\begin{equation}
\label{Gnu2}
G(u) = \lim_{n \rightarrow \infty} G_n(u) = \exp\left[ -\frac{(-u)^b}{\beta(a,b)b} \right],
\end{equation}
and thus we find the (reversed) Weibull or Type III distribution (with the term ``Type III'' referring to standard order statistics terminology) as the limiting law $G(u)$ of $u = a_n(\pi_{(n)} - b_n)$, with the respective choices of scaling and shift parameters $a_n = n^{1/b}$ and $b_n = 1$ as given above.

We refer to Appendix \ref{sec:moments} for using the limiting form of \eqref{Gnu2} to derive the first moment of each of the ordered random variables $\pi_{(j)}$, indexed from smallest to largest.  Appendix \ref{sec:moments} shows that these take the form of generalized Gamma variates, and their moment characterization will be an initial indication of the order statistic structure. We can note directly from Appendix \ref{sec:moments} that
$$
\operatorname{\mathbb{E}} \pi_{(j)} = \frac{ \operatorname{\mathbb{E}} u_{(j)}}{a_n} + b_n = 1 - \frac{\operatorname{\mathbb{E}}(- u_{(j)})}{n^{1/b}} = 1 - \frac{\Gamma(1/b)}{\beta(n-j+1,1/b)} \left(\frac{(\beta(a,b)b)}{n}\right)^{1/b}.
$$
Finally, we may appeal to the relation $\beta(n-j+1,1/b) \sim \Gamma(1/b)(n-j+1)^{-1/b}$ for $n-j+1$ large and $1/b$ fixed, whereupon, reverting to the notation that $\pi_{(k)}$ denotes the $k$th \emph{largest} weight, we obtain the following approximation:
\begin{equation}
\label{expectation}
\operatorname{\mathbb{E}} \pi_{(k)} \sim 1 - \left(\beta(a,b)b \, \frac{k}{n}\right)^{1/b}.
\end{equation}
This yields a simple and elegant form of the expected value of each order statistic, where the decay of the sequence is governed by increasing $k$.

\subsection{General result for $\operatorname{Beta}(a,b)$ with $a,b >0$, and results for minimum degrees}

Note that by properties of the Beta integral, $I_x(a,b) = 1 - I_{1-x}(b,a)$, and observe~\cite[Chs.~15, 26]{abramowitz1965handbook} that $I_{1-x}(b,a)$ admits the following series expansion for all real-valued $a,b > 0$, where we interpret $\Gamma(1-a+m)/\Gamma(1-a)$ as the rising factorial $(1-a)_m$:
\begin{align*}
I_{1-x}(b,a) & = \frac{(1-x)^b}{\beta(a,b)} \sum_{m=0}^\infty \frac{\Gamma(1-a+m)}{\Gamma(1-a)m!(b+m)}(1-x)^m
\\ & = \frac{(1-x)^b}{\beta(a,b)} \left(\frac{1}{b} + \sum_{m=1}^\infty \frac{\Gamma(1-a+m)}{\Gamma(1-a)m!(b+m)}(1-x)^m \right).
\end{align*}
Expanding $x$ near unity in $-u$ as before via $x = 1 + u/n^{1/b}$, we see that
$$
\left[I_x(a,b)\right]^n = \left[1 - \frac{(-u)^b/n}{\beta(a,b)b}\left(1 + b\sum_{m=1}^\infty \frac{\Gamma(1-a+m)}{\Gamma(1-a)m!(b+m)}[(-u)^m/n^{m/b}] \right) \right]^n, \quad a,b > 0,
$$
and thus by the same limiting argument as before, our earlier result holds for all $a,b > 0$.  Moreover, we may apply the same logic to look at the minimum order statistic by expanding $x$ near zero in $u$ via $x = u/n^{1/b}$, obtaining the Type II limiting law (again with the term ``Type II'' referring to standard order statistics terminology) as $G(u) = \exp[ -(\beta(a,b)a)^{-1} u^a ]$, implying that $\operatorname{\mathbb{E}} \pi_{(j)} \sim (\beta(a,b)a \cdot j/n)^{1/a}$ for the $j$th smallest order statistic $\pi_{(j)}$.

\subsection{Variance expression}

As a final note, observing that $\operatorname{\mathbb{E}}[(- u_{(k)})^m ] = [\beta(a,b)b]^m \Gamma(k + m/b) / \Gamma(k)$ for the $k$th largest order statistic, we obtain
\begin{align*}
\operatorname{Var}\left(- u_{(k)} \right) & = \frac{(\beta(a,b)b)^2 \, \Gamma(k + 2/b)}{\Gamma(k)} - \left( \frac{\beta(a,b)b \, \Gamma(k + 1/b)}{\Gamma(k)} \right)^2
\\ &  = \left[ \operatorname{\mathbb{E}}\left( - u_{(k)} \right) \right]^2 \left[ \frac{\beta(k,1/b)}{\beta(k + 1/b,1/b)} - 1 \right].
\end{align*}
Thus
\begin{align*}
\operatorname{Var}\left(\pi_{(k)} \right) & = \frac{1}{a_n^2}\left[ \operatorname{\mathbb{E}}\left( - u_{(k)} \right) \right]^2 \left[ \frac{\beta(k,1/b)}{\beta(k + 1/b,1/b)} - 1 \right]
\sim \left[ 1 - \operatorname{\mathbb{E}} \pi_{(k)} \right]^2 \left[ \left(\frac{k + 1/b}{k}\right)^{1/b} - 1 \right].
\end{align*}
We see directly that (not unreasonably) the variances of the order statistics decrease as their values approach more and more closely to 1.

\section{Simulation study}
\label{sec:simStudy}

To verify the order statistics properties derived above, we undertook a small simulation study based on the log-linear model of~\cite{Chung2002pnas}.  Edges in this model are independent Bernoulli trials with success probabilities $p_{ij}$ for all $1 \leq i \leq j \leq n$. A nonnegative weight $w_i$ is associated to each node $i$, with $w_1 / n, \ldots w_n/ n \sim \operatorname{Beta}(a,b)$, and we then set $p_{ij} = w_i w_j / \|\bm{w}\|_1$, where $\|\bm{w}\|_1 = \sum_{k=1}^n w_k$.  When $w_i^2 \leq \|\bm{w}\|_1$ for all $i$, it follows that the expected degree $\operatorname{\mathbb{E}}(d_i)$ of the $i$th node is equal to $w_i$, and thus the unnormalized weights $w_1, \ldots w_n$ can be interpreted as expected degrees.

\begin{figure}[!t]
\begin{center}
\subfigure[$\pi \sim \operatorname{Beta}(1,9)$; $\operatorname{\mathbb{E}}(\pi) = 1/10$]{\includegraphics[width=0.495\columnwidth, trim = 25 20 25 15, clip]{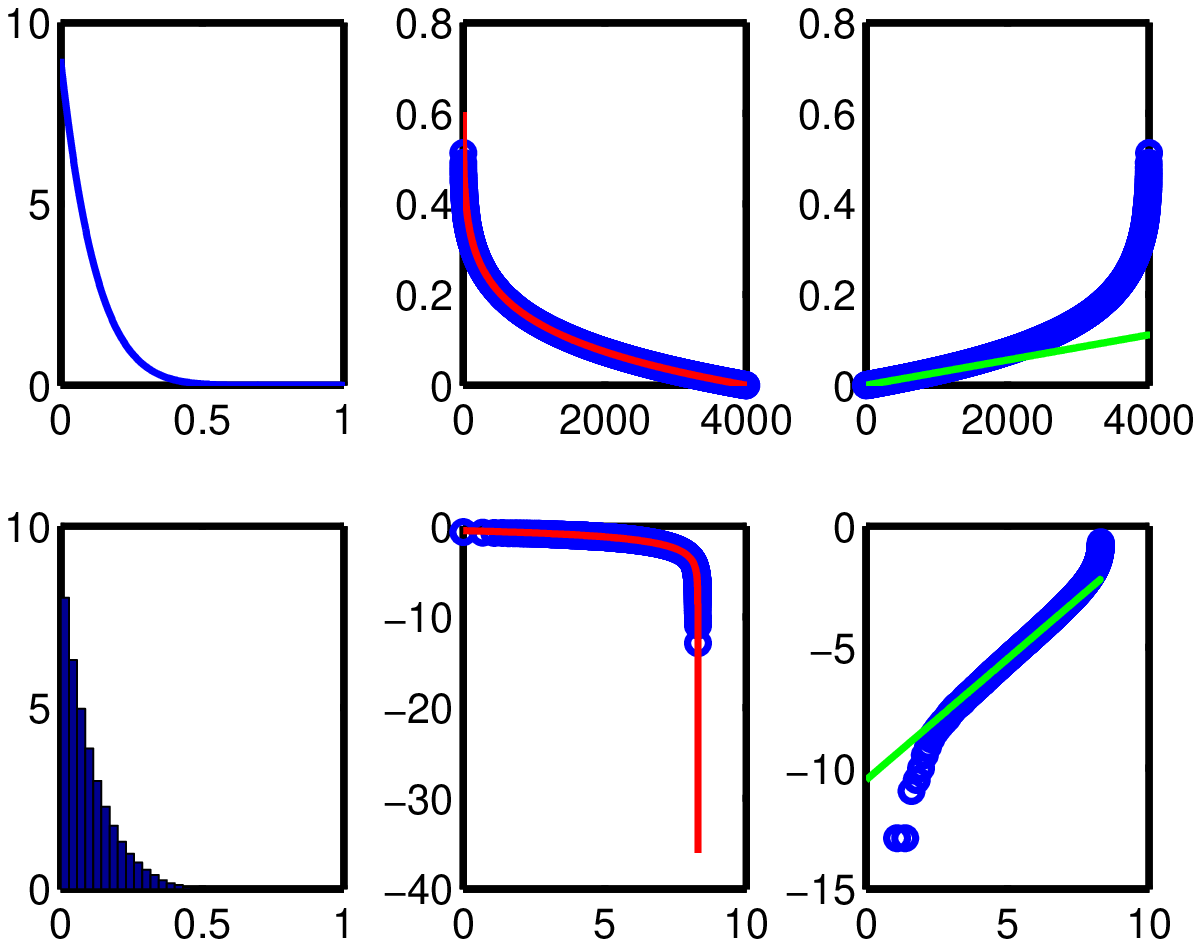}}
\subfigure[$\pi \sim \operatorname{Beta}(2,4)$; $\operatorname{\mathbb{E}}(\pi) = 1/3$]{\includegraphics[width=0.49\columnwidth, trim = 25 20 25 15, clip]{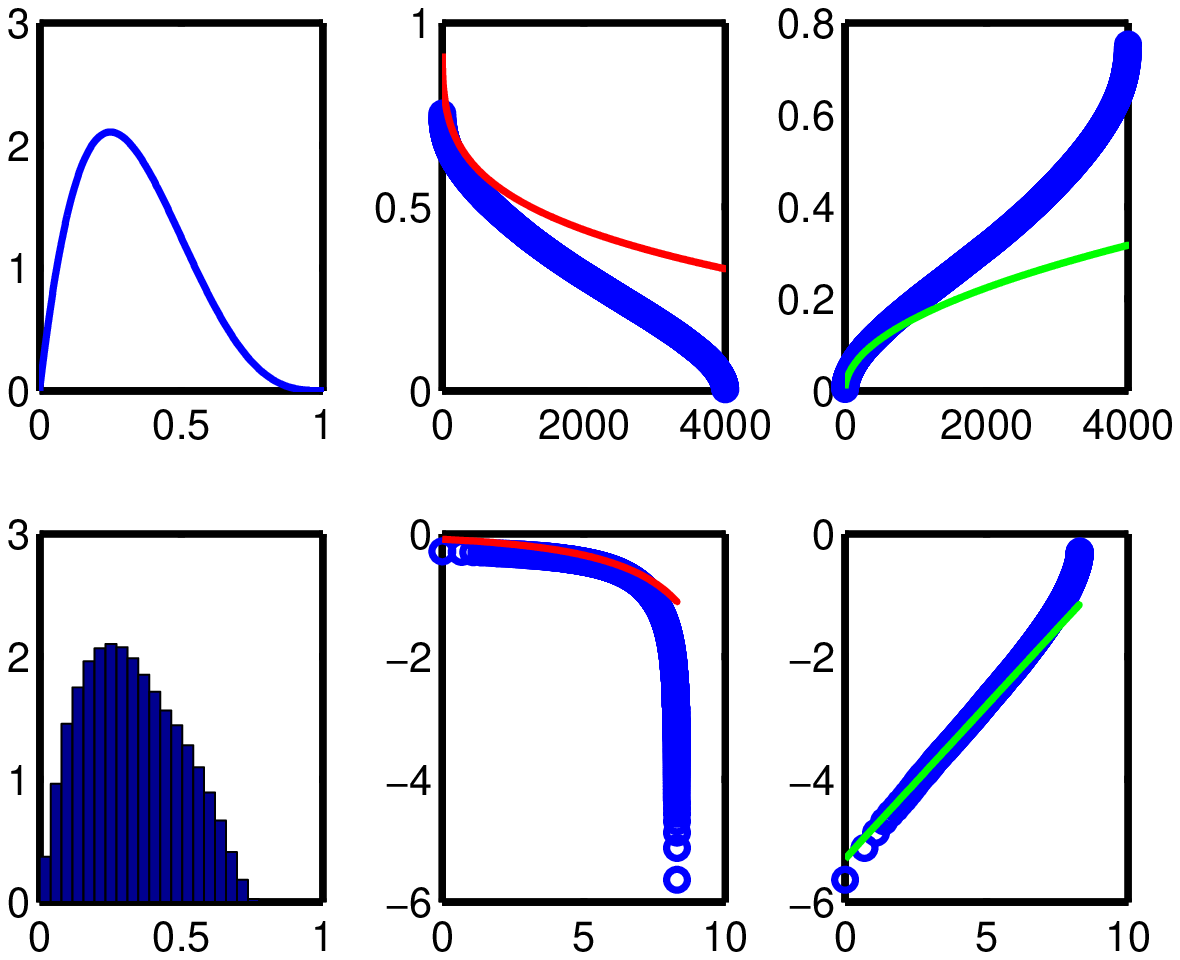}}
\end{center}
\caption{\label{fig:BetaSims} Examples showing simulated network degrees for graphs with $n=4000$ nodes, along with their empirical and predicted order statistics averaged over 100 trials, for the settings (a) $w_i/n = \pi_i \sim \operatorname{Beta}(1,9)$ and (b) $w_i/n = \pi_i \sim \operatorname{Beta}(2,4)$ according to the model of Section~\ref{sec:simStudy}.  The leftmost column in each case shows the exact and empirical normalized degree distributions; the middle column the maximal order statistics on linear and log-log scales; and the rightmost column the minimal order statistics on linear and log-log scales.}
\end{figure}

Figure~\ref{fig:BetaSims}
overleaf shows two examples of network degrees simulated under this model, with average degrees respectively $n/10$ and $n/3$.  From these examples it can be seen that the approximate maximal and minimal order statistic expectations agree well with empirical averages over parts of their respective ranges (recalling that these results hold for $n$ large and $k,j$ fixed).  In Fig.~\ref{fig:BetaSims}(a) with $w_i/n \sim \operatorname{Beta}(1,9)$, we confirm that the maximal order statistic expression holds throughout the sequence.  In this case $\beta(a,b)b = 1$, further simplifying this expression to $\operatorname{\mathbb{E}} \pi_{(k)} \asymp 1 - (k/n)^{1/b}$.  In Fig.~\ref{fig:BetaSims}(b) with $w_i/n \sim \operatorname{Beta}(2,4)$, we also see reasonable agreement between expected and empirical quantities over the initial portions of their respective ranges.

\section{Discussion}\label{sec:discussion}

In this article we have considered normalized degree sequences generated from the $\operatorname{Beta}(a,b)$ family of random variables. This led us to be able to derive forms for the expected extrema of such random variables, as well as their variances, in order to better understand their sampling characteristics in the context of random graphs. As such, we have been able to characterize the \emph{expected} extreme behavior seen in network degrees that exhibit different characteristics due only to random variability, and not to difference in model parameters \emph{per se}.  In contrast to this notion of structure due only to ordering, we note that a popular model for the extreme \emph{expected} degrees is the power law \cite{chung2003spectra}:
\begin{eqnarray}
\label{power-law}
(\operatorname{\mathbb{E}}\pi)_{(k)} =\frac{c}{(s+k)^\gamma},\quad c,\,s,\,\gamma\, \in {\mathbb{R}}^+,\,k \in {\mathbb{N}}.
\end{eqnarray}
Here $\gamma$ is the power law exponent, $c$ is an overall scaling, and $s$ is a shift parameter that in combination with $c$ controls the maximal and average degree values.  If we compare \eqref{power-law} with the expected values of the ordered normalized degrees under our model, then from \eqref{expectation} we note
\begin{equation}
\operatorname{\mathbb{E}} \pi_{(k)} \sim 1 - \left(\beta(a,b)b \, \frac{k}{n}\right)^{1/b} \sim \frac{1}{1+\left(\beta(a,b)b \, \frac{k}{n}\right)^{1/b}}=\frac{\left[\frac{n}{\beta(a,b)b}\right]^{1/b}}{\left[\frac{n}{\beta(a,b)b}\right]^{1/b}+k^{1/b}},
\; k\ll n.
\label{taylor-order}
\end{equation}
Comparing \eqref{power-law} to \eqref{taylor-order}, we see both similarities and differences in structure. Both expressions are decreasing in increasing $k$, and show a smooth decay in $k$. The distinction between the two is the importance of the order of the shift parameter in each case. If we apply the form of \eqref{taylor-order}, the initial decay in $k$ will be far less dramatic than that of \eqref{power-law}, as the shift in \eqref{taylor-order} will by necessity be large and harder to dislodge by the power in $k$.  Whether \eqref{power-law} or \eqref{taylor-order} more realistically models a given network will depend on the application from which the network arose; furthermore, from this comparison we see that improper selection of the number of degrees used to estimate the power law exponent can potentially lead to erroneous conclusions, especially if the magnitudes of $s$ and $c$ are not investigated when fitting the model of~\eqref{power-law} to an observed network.

Having compared our results to a power law, a number of other conclusions are apparent. Order statistics {\em will} create an apparent hierarchy in the normalized degrees. This hierarchy is structured. It takes the simplest form if the parameters of the Beta distribution we have considered take special values; if not, decay is still monotone and predictable. A palatable feature of normalized degrees as we have introduced them is that they are easily comparable across varying network sizes, and scale naturally; thus, with varying sample sizes, the same scaling structure is achieved.

Our simulation study has shown the practical performance of our results for different choices of normalized degree distributions.  We find good agreement with our theoretical results in the respective tails---exactly as predicted by order statistics theory. We know that our limiting results hold only for indices near the end of the range; i.e., near the maximum or minimum of the range of the random variables under consideration.  If we instead wanted to consider degrees taking values nearer to the center of the distribution, then it would be necessary to apply other theoretical constructions not based on order statistics.  In this setting, results would depend differently on the particular choice of distribution employed.

\subsubsection*{Acknowledgments}

Work supported in part by the US Army Research Office under PECASE Award W911NF-09-1-0555 and by the US Office of Naval Research under MURI Award 58153-MA-MUR, and by the UK EPSRC under a Mathematical Sciences Leadership Fellowship EP/I005250/1 and the UK Royal Society under a Wolfson Research Merit Award.

\appendix

\section{Simplifying the limiting form of $G_n(u)$}\label{sec:simp}

We shall simplify the form of $G_n(u)$.  More formally we write  $G_n(u)$ as
\begin{equation}
\label{eq:GnU}
G_n(u) = \left[1 - \sum_{i=0}^{a-1} \binom{a+b-1}{i} \left(1 + \frac{u}{n^{1/b}}
\right)^i \left( \frac{-u}{n^{1/b}} \right)^{a+b-1-i} \right]^n, \quad a,b
\in \mathbb{N}.
\end{equation}
Our goal will be to bound the sum near its largest term in the polynomial, which
will dominate the others.  This term is given by $[\beta(a,b)b]^{-1} [1
+ u/n^{1/b} ]^{a-1} [ -u / n^{1/b} ]^{b}$, and, factoring
out this term, we see that the sum in~\eqref{eq:GnU} is given by
\begin{multline*}
\frac{1}{\beta(a,b)b} \left(1 + \frac{u}{n^{1/b}} \right)^{a-1} \frac{(-u)^{b}}{n}
\left\{1 + \beta(a,b)b \sum_{i=0}^{a-2} \binom{a+b-1}{i} \left(-\frac{u}{n^{1/b}+u}
 \right)^{a-1-i}  \right\}.
\end{multline*}

We may bound this sum via the binomial expansion of $[1 - u/(n^{1/b}+u)]^{a-2}$
as follows:
\begin{align*}
\sum_{i=0}^{a-2} \binom{a\!+\!b\!-\!1}{i} \left(\frac{-u}{n^{1/b}\!+\!u}
 \right)^{a-1-i} \!\! & \leq \frac{\Gamma(a\!+\!b)}{\Gamma(a\!-\!1)} \frac{\Gamma(1)}{\Gamma(b\!+\!2)}
\left(\frac{-u}{n^{1/b}\!+\!u}  \right) \sum_{i=0}^{a-2} \binom{a\!-\!2}{i}
\left(\frac{-u}{n^{1/b}\!+\!u}  \right)^{a-2-i}
\\ & = \frac{1}{\beta(a,b)b} \left(\frac{a-1}{b+1}\right)  \frac{(-u) / n^{1/b}}{\left(1
+ u/n^{1/b}\right)^{a-1}}.
\end{align*}
Thus $G_n(u)$ for $a,b \in \mathbb{N}$ is given by
$$
G_n(u) = \left[1 - \frac{1}{\beta(a,b)b} \left(1 + \frac{u}{n^{1/b}} \right)^{a-1}
\left( \frac{-u}{n^{1/b}} \right)^{b} \left\{ 1 + f_{ab}(u,n) \right\} \right]^n,
$$
where $| f_{ab}(u,n) | \leq   \left[(a-1)/(b+1)\right] (-u / n^{1/b}) \left(1
+ u/n^{1/b}\right)^{1-a} \mathbb{I}(a \geq 2)$. This will help us derive the limiting form of $G_n(u)$.

\section{Calculating moments of the $j$th smallest order statistic}\label{sec:moments}

We now switch to standard notation for the Weibull to obtain moments of this
limiting form.  We are interested in the maximum order statistic $j=n$, as
well as others for fixed $j$ not depending on $n$:
$$
\operatorname{\mathbb{E}} \pi_{(j)} = \frac{ \operatorname{\mathbb{E}} u_{(j)}}{a_n}
+ b_n = 1 - \frac{\operatorname{\mathbb{E}}(- u_{(j)})}{n^{1/b}}.
$$
To obtain these results we use Theorem~1.3.3 of~\cite{Leadbetter}, which
states that for order statistics of a sequence of independent and identically
distributed random variables $\{\pi_i\}_{i=1}^n$, if for some sequences
$\{a_n > 0\}, \{b_n\}$ of real constants, $a_n(\pi_{(n)} - b_n)$ converges
in distribution to a random variable with nondegenerate distribution function,
then for any fixed positive integer $j$ we have that
$$
\operatorname{\mathbb{P}}\left( a_n(\pi_{(j)} - b_n) \leq x \right) \rightarrow
G(x) \sum_{i=0}^{n-j} \frac{\left(-\log G(x)\right)^i}{i!},
$$
with the same $a_n,b_n$.  We remark that the formula is for the $j$th smallest, so that the maximum $j=n$ will involve only one term---the sum in this case terminating after the zeroth term.

To employ this result we must evaluate the above expression for the Type
III law in question.  The corresponding probability density function (PDF)
is the $\operatorname{Weibull}(b,\lambda)$ distribution, where $ b \in
\mathbb{N}, \lambda = (\beta(a,b)b)^{1/b} $ in our previous
notation:
\begin{align*}
F(x;b,\lambda) & = 1 - \exp[ -(x/\lambda)^b ], \quad x \geq 0;
\\ f(x;b,\lambda) & = \frac{b}{\lambda} \left( \frac{x}{\lambda} \right)^{b-1}
\exp[ -(x/\lambda)^b ], \quad x \geq 0;
\\ \operatorname{\mathbb{E}} X^m & = \lambda^m\Gamma\left(1 + \frac{m}{b}
\right).
\end{align*}

Now observe that we may express the corresponding limiting law of $u_{(j)}$, the $j$th smallest variable, as:
$$
G_{u_{(j)}}(u) = G(u) \sum_{i=0}^{n-j} \frac{\left(-\log G(u)\right)^i}{i!},
$$
where $G(u) = \exp\left[ -(-u)^b / (\beta(a,b)b) \right] = \exp[ -(-u/\lambda)^b
]$, as
$$
G_{u_{(j)}}(u) = \exp[ -(-u/\lambda)^b ] \sum_{i=0}^{n-j}  \frac{(-u/\lambda)^{ib}}{i!}.
$$
It follows that the limiting PDF $g_{u_{(j)}}(u)$ is given by the generalized
Gamma distribution:
\begin{align*}
g_{u_{(j)}}(u) & = \frac{1}{\Gamma(n-j+1)} \frac{b}{\lambda} \left( \frac{-u}{\lambda}
\right)^{(n-j+1)b-1} \exp[ -(-u/\lambda)^b ], \quad u \leq 0;
\\ \operatorname{\mathbb{E}}\left( [- u_{(j)}]^m \right) & = \lambda^m \frac{\Gamma(n-j+1
+ m/b)}{\Gamma(n-j+1)} = \frac{\lambda^m \, \Gamma(m/k)}{\beta(n-j+1,m/b)},
\end{align*}
where the latter expression follows as the moments of the generalized Gamma are known.

\subsubsection*{References}
\vspace{-2\baselineskip}%
\small{%
\bibliographystyle{unsrt}
\bibliography{networks}
}

\end{document}